\documentclass[aps]{revtex4}

\usepackage{amsmath, latexsym, amssymb, bm, amsfonts,
mathbbol,graphicx,subfigure}

\renewcommand{\v}[1]{\mathbf{#1}}
\newcommand{\na}[0]{\nabla}

\renewcommand{\Re}[0]{\mathfrak{Re}\!~}
\renewcommand{\Im}[0]{\mathfrak{Im}\!~}
\newcommand{\ra}[0]{\rangle}
\newcommand{\la}[0]{\langle}

\newcommand{\wh}[1]{\widehat{#1}}
\newenvironment{mat}[1]{\begin{array}{@{}*{#1}{r@{}l}@{}}}{\end{array}}
\DeclareMathOperator{\tr}{tr}
\DeclareMathOperator{\diag}{diag}

\begin{document}

\title{Determination of the characteristic directions of lossless
linear optical elements}

\author{Hanno Hammer}
\email{hh@sensordynamics.cc}
\affiliation{SensorDynamics AG  \\
Schlo\ss\ Eybesfeld 1e \\
A--8403 Graz-Lebring, Austria%
}
\date{\today}
%
\begin{abstract}
We show that the problem of finding the primary and secondary characteristic
directions of a linear lossless optical element can be formulated in terms of an
eigenvalue problem related to the unimodular factor of the transfer matrix of the
optical device. This formulation makes any actual computation of the characteristic
directions amenable to pre-implemented numerical routines, thereby facilitating the
decomposition of the transfer matrix into equivalent linear retarders and rotators
according to the related {\it Poincar\'e equivalence theorem}. We explain in detail
how this issue arises in the context of stress analysis based on integrated
photoelasticity or hybrid methods combining photoelastic measurements with analytical
stress models and/or numerical Finite-Element computations for the stress tensor
field. Furthermore we show how our results can be applied when algorithms for the
reconstruction of the dielectric tensor in the interior of a photoelastic model
(dielectric tensor imaging) are tested for their stability against noise in the
measurement data. For the sake of completeness we provide a brief derivation of the
basic equations governing integrated photoelasticity.

\vspace{1em}
{\bf Keywords:} Equivalent optical model. Poincar\'e Equivalence Theorem. Integrated
Photoelasticity. Dielectric tensor imaging. Stability of reconstruction algorithms.
\end{abstract}
\maketitle


\section{Introduction}

Passive linear optical elements are ubiquitous in the study of interactions between
matter and polarized classical~\cite{BornWolf,Longhurst,Ditchburn,Hecht,Guenther} or
quantum light~\cite{Leonhardt-ROP66}. While polarizers attenuate one of two
distinguished orthogonal polarization forms, linear retarders and rotators alter the
state of polarization while preserving the flow of light energy through the
device. The latter two belong to the class of non-absorbing linear elements, where
the term ``linear'' refers to the fact that the action of such a device on the state
of polarization can be conveniently described by a unitary two-by-two {\it transfer
matrix}. In contrast, a polarizer would be represented by a Hermitean matrix. This
description of linear optical elements in terms of Hermitean and unitary matrices is
called the {\it Jones formalism}~\cite{Jones1941a, Jones1941b, Jones1941c,
Jones1942a, Jones1947a, Jones1947b, Jones1948a, Jones1956a, Jones1956b}.

In his treatise on classical light~\cite{Poincare1892}, Poincar\'e found that any
non-absorbing passive linear optical element could be decomposed into basic linear
retarders and rotators. By linear retarder we mean a homogeneously anisotropic
device, such as a piece of appropriately cut crystal, which possesses two preferred
axes, called the ``fast'' and ``slow'' axis, which are perpendicular to each other
and differ in the phase velocity of component waves linearly polarized along the
distinguished directions; as a consequence, light possessing a general elliptic
polarization state will accumulate a relative phase retardation between the two
components and thus change its polarization form. A rotator, on the other hand,
changes the plane of linearly polarized light by a specified angle; it can be shown
easily that this effect is due to a phase retardation between the two orthogonal
components of {\it circular} polarization.  Accordingly, a linear retarder is
determined by specification of, e.g., the angle of the fast axis, and the relative
phase retardation; while a single rotation angle is sufficient to specify a
rotator. This decomposition of a general non-absorbing optical element into retarders
and rotators is called the {\it Poincar\'e equivalence theorem} (see
Ref.~\cite{HammerJModOpt2004a} for a recent account).

Such a decomposition proves very useful whenever the determination of the internal
stress tensor of a transparent medium by means of non-destructive methods is
desired. This issue arises in the following context: material scientists and
engineers wish to gain insight into the stress and strain fields in the interior of a
loaded specimen. Amongst the many different approaches, three methods are most often
used: (1) an educated guess at the analytic mathematical form of the internal stress
or strain tensor is made; (2) the external load acting on the specimen is
systematically approximated by forces applying on finitely many points on the {\it
surface} of the object; and the resulting stress in the interior is then computed
using numerical schemes generically called {\it Finite Element Methods}
(FEM)~\cite{CookFEM,HolandBellFEM,DankertDankert}; (3) the phenomenon of {\it
photoelasticity} is utilized to gain insight into the internal stress
distribution. The term photoelasticity refers to the fact that some transparent
materials, typically resins or glasses, become birefringent when under external load,
while being optically isotropic in the unloaded state. Within certain limits, the
relation between the dielectric tensor $\epsilon_{ij}$ and the stress tensor
$\sigma_{ij}$ is linear, and is generically called a {\it stress-optical law}. Its
typical form has been given long ago by Maxwell~\cite{MaxwellStrOpt},
\begin{equation}
\label{stress1}
 \epsilon_{ij} = \epsilon \, \delta_{ij} + C_1 \, \sigma_{ij} + C_2 \, \tr{\sigma} \,
 \delta_{ij} \quad,
\end{equation}
where $C_1, C_2$ are called stress-optical constants. This one-to-one relation
between dielectric tensor and stress tensor suggests that one builds a model of an
industrial component from an appropriate material (today, plastics are usually used)
and loads the model just as the real object. The model is then subjected to heat, so
as to loosen, and rearrange, the molecular bonds in the material; and subsequently
cooled down, or ``stress-frozen'', upon which the stress pattern remains locked
inside the material~\cite{Frocht1,Frocht2,CokerFilon}. To determine these patterns,
destructive and non-destructive methods are available.

A particular example of non-destructive evaluation has been termed ``Integrated
Photoelasticity''~\cite{Aben1966a}. Here, polarized light is sent through the
specimen at many different angles, and the change in polarization form is registered
for each light ray (pixelwise), utilizing appropriate combinations of polariscopes
and digital cameras. To each ray passing through a specified point, and along a
specified direction, a unitary transfer matrix $U$ can be assigned, which describes
how the polarization form changes along the given ray. In this way we obtain a map
from the set of all (reasonably smooth) dielectric tensor, or stress tensor, fields
in the interior of the specimen, to the collection of transfer matrices, gathered for
all necessary points and directions. In the theory of Inverse Problems and
Mathematical Tomography
\cite{Kirsch1996,Denisov1999,Bukhgeim1999,HelgasonRadon,Natterer,EHN}, this map is
commonly called the {\it forward problem}. The associated {\it inverse problem}
consists of reconstructing the interior dielectric tensor, or stress tensor, from a
given collection of transfer matrices.

The transfer matrices $U$ are not observable directly. Rather, they are known
functions of the global phase of the light beam, and three so-called ``characteristic
parameters''~\cite{Aben1966a}, two of which may be regarded as polarization
directions, while the third one has the meaning of a ``characteristic phase
retardation''. Their operational meaning is as follows: the {\it primary
characteristic directions} determine those planes of linear polarization at the entry
into the medium for which the state of polarization of the emergent beam is again
linear. The {\it secondary characteristic directions} determine the planes of linear
polarization of the emerging light, if the incident light was linearly polarized in
the primary characteristic directions; in general, they differ from the primary
ones. There are always two orthogonal primary and two orthogonal secondary
characteristic directions such that light which is linearly polarized along the two
primary directions travels with different phase velocities. As a consequence, both
waves emerge with a phase difference---the characteristic phase retardation. The {\it
characteristic parameters}~\cite{Aben1966a} of a linear lossless device are usually
taken to be simple linear combinations of the angles specifying one of the primary
characteristic directions, and the angle of the associated secondary characteristic
direction; and the characteristic phase retardation (see
section~\ref{PoincareEqTheorem}).

Traditionally, the forward problem in integrated photoelasticity is formulated
directly in terms of the bulk stress tensor; however, more recent
attempts~\cite{HammerEA-JOSAA-2005a} have shown that it may be advantageous to first
determine the dielectric tensor inside the model by tomographic means, after which
the stress tensor can be computed via the stress-optical law in
eq.~(\ref{stress1}). The solution to the inverse problem for the two-dimensional (2D)
case, i.e., when the specimen is a thin slab, and the stress tensor inside possesses
only two principal directions, is known long since~\cite{Frocht1,Frocht2,%
CokerFilon,TheocarisEA,Ditchburn,Aben1966a,Aben1979,AbenGuillemet,AbenEA1989a}.  The
general solution to the three-dimensional (3D) inverse problem is not known to date,
although it has been shown recently~\cite{HammerEA-JOSAA-2005a} that, in the limit of
weak optical anisotropy, the 3D inverse problem can be mathematically reduced to six
independent 2D inverse problems.

Being one of the oldest methods of experimental stress analysis, photoelasticity has
been somewhat overshadowed by FEM methods over the past two or three decades, but has
seen a recent revival with applications in silicon wafer stress analysis, rapid
prototyping, fiber optic sensor development, and image
processing~\cite{MasundiNet}. As photoelasticity can be regarded as a natural
complement to FEM numerics, hybrid methods attempting at combining both approaches
are becoming popular. It then becomes a viable task to compare FEM results with those
obtained by photoelastic tomography.

\begin{figure}[h]
\begin{minipage}[b]{1\textwidth}
\subfigure[
   \label{fig1a} ]{ \includegraphics[width=.2\textwidth]%
{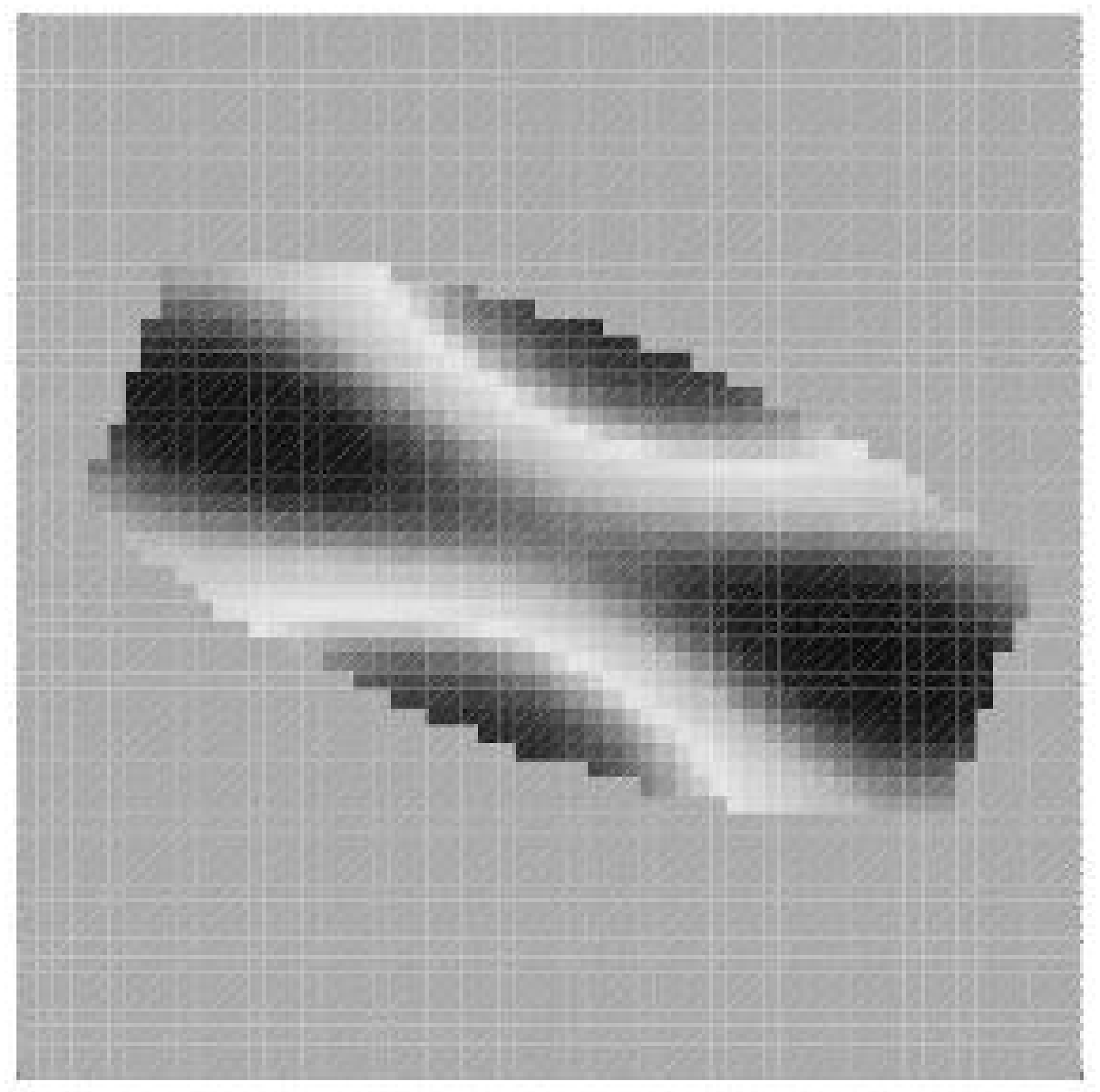}}
\hspace{3em}
\subfigure[
   \label{fig1b} ]{ \includegraphics[width=.2\textwidth]%
{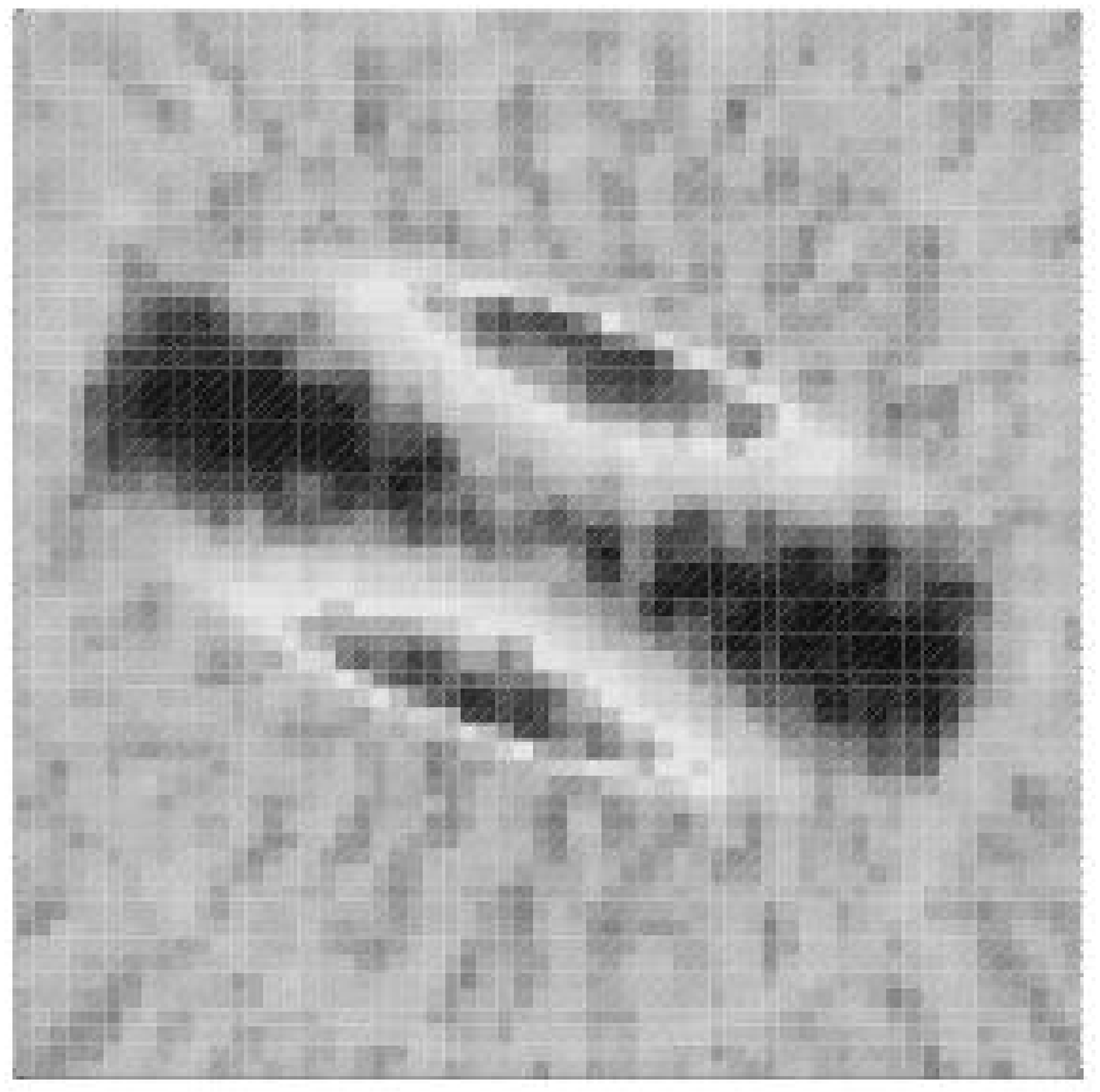}}
\end{minipage}
\caption{A cylinder is subject to axial load; the resulting deformation is modelled
as a simulated displacement field on the surface and in the interior. Hooke's law
gives the relation between the resulting strain and stress tensors; from the
stress-optical law, eq.~(\ref{stress1}), the dielectric tensor is then obtained. This
is an example of an artificial stress model giving rise to a simulated dielectric
tensor which can be used to test photoelastic reconstruction algorithms as follows: a
plane intersecting the cylinder at an angle of $22^{\circ}$ relative to the symmetry
axis of the cylinder is chosen in the figure. The unit normal vector to this plane is
$\bm{\eta}$. The component of the dielectric tensor in the direction of $\bm{\eta}$
is $\epsilon_{\eta \eta} = \epsilon( \bm{\eta}, \bm{\eta})$. In subfigure (a), its
relative deviation $(\epsilon_{\eta \eta} - \epsilon)/ \epsilon = A_{\eta \eta}$ from
the scalar isotropic value $\epsilon$ is plotted. In subfigure (b), the simulated
dielectric tensor $\epsilon_{ij}$ is used to compute a collection of transfer
matrices $U$ via the forward problem, eq.~(\ref{forward}). The transfer matrices are
then decomposed according to the equivalent optical model, using the method
introduced in section~\ref{CharaDirections}; subsequently, {\it Gaussian noise} is
imposed on the characteristic parameters, eq.~(\ref{actrot5}), thus simulating
measurement errors. The noisy parameters then recombine, again via the equivalent
optical model, to give new, ``noisy'', transfer matrices, from which the ``noisy''
dielectric tensor can be reconstructed via a filtered back-projection
algorithm~\cite{Natterer,EHN}. The result of this reconstruction, for $60 \times 60$
pixels and $90$ scans around the object, is plotted in subfigure~(b). It is seen that
the noise in the characteristic parameters, and the noise in the global phase $\phi$,
eq.~(\ref{action101}), give rise to a halo-like pattern around the object, in
addition to the speckled appearance of the reconstruction. -- The comparison of
Figures (a) and (b) provides a visual example of how our decomposition method can be
used in order to test the stability of a reconstruction algorithm (here: filtered
back-projection) against measurement errors. \label{fig1}}
\end{figure}

These considerations describe the context in which the method of determining the
characteristic directions of a linear optical element, as given in this paper, is
expected to prove useful. Specifically, there are four applications of our results in
integrated photoelasticity and associated hybrid methods:
\begin{enumerate}
\item \label{item1}
Testing the validity of analytic stress models: Here, we start with an educated guess
about the analytic form of the bulk stress tensor field; then compute the associated
dielectric tensor via the stress-optical law in eq.~(\ref{stress1}); and use this to
compute the transfer matrices of the forward problem numerically, using
eq.~(\ref{forward}) in section~\ref{Equations} below. In order to facilitate a
comparison with the characteristic parameters obtained from a direct photoelastic
evaluation of a stress-frozen model, a decomposition of the numerically obtained
transfer matrices into equivalent retarders and rotators must be performed for each
light ray. Such a decomposition has to evaluate the characteristic directions of a
given transfer matrix $U$ first; subsequently, the associated characteristic phase
retardation follows automatically, as will be shown in
section~\ref{CharaDirections}. It is here where our method of determining the
characteristic directions is needed.
\item
Testing the validity of a FEM-based calculation: instead of an analytic stress model
we might want to compare a bulk stress tensor field evaluated numerically using FEM,
with photoelastic data. As in the last item, the stress tensor leads to a dielectric
tensor and subsequently to a collection of numerically computed transfer matrices via
the forward problem, eq.~(\ref{forward}). Again, the comparison with actual
measurements made on a photoelastic specimen requires the decomposition of these
transfer matrices according to the equivalent optical model.
\item
Iterative solutions of the inverse problem in integrated photoelasticity: the
integral equation~(\ref{forward}) determining the transfer matrices in terms of the
dielectric tensor field is non-linear; this accounts for the fact that the full
solution to the general 3D inverse problem of determining the dielectric tensor in
terms of the transfer matrices is not yet known. Iterative numerical schemes to solve
eq.~(\ref{forward}) for $\epsilon_{ij}$ may be conceived. At each stage of iteration,
the intermediate result for the dielectric tensor can be used to compute a collection
of associated transfer matrices, whose characteristic parameters, in turn, may be
compared with actual photoelastic data in order to check the convergence of the
iterative algorithm. At this last stage we again need the decomposition of a given
transfer matrix in terms of the equivalent optical model.
\item
Stability of inversion algorithms on noisy data: the reliability of a reconstruction
depends on how stable the algorithm is with respect to finite errors in measurement
data. This stability can be tested on artificial analytic stress models, as follows:
the stress model can be converted into a collection of transfer matrices as explained
in item~\ref{item1}.) above. We then decompose each $U$ into characteristic
parameters ; but instead of feeding these data directly into the inversion we
subsequently add some numerical noise to the characteristic parameters, thus
simulating errors in the measuring apparatus. These modified data are then fed into
the proposed algorithm to check how far the new reconstruction deviates from the one
without noise in the data. A visual example of this procedure is given in
Fig.~\ref{fig1} for a standard reconstruction algorithm (filtered
back-projection~\cite{Natterer}).
\end{enumerate}

In Ref.~\cite{HammerJModOpt2004a} we have given a detailed account of the Poincar\'e
equivalence theorem and its application to construct the equivalent optical model. In
that paper, the decomposition of a given transfer matrix $U$ into retarders and
rotators was accomplished in a somewhat pedestrian fashion, by parameterising the
(three-dimensional) manifold of $SU(2)$ matrices in a suitable way, and then applying
elementary trigonometric methods. While mathematically correct, the parameterisation
of the $SU(2)$ manifold as used in Ref.~\cite{HammerJModOpt2004a} is inconvenient for
computer-based algorithms in that subsets of the boundary of the parameter space of
this manifold must be identified in order to obtain a one-to-one relation between
parameters and matrices. This non-trivial topology introduces complications when used
in the numerical schemes outlined above. In the present paper we accomplish the same
decomposition in a more elegant way: we show how to reformulate the problem of
finding the ``characteristic'' data specifying the equivalent optical model for a
given transfer matrix in terms of an eigenvalue problem associated with the
unimodular factor of this matrix. The new method is therefore more suitable for an
actual numerical determination of the characteristic data, as we can immediately make
use of pre-implemented numerical routines for eigenvalue problems.

The plan of the paper is as follows: in section~\ref{Equations} we briefly discuss
the derivation, and physical content, of the basic equations governing integrated
photoelasticity, leading to the integral equation~(\ref{forward}) for the transfer
matrices in terms of the dielectric tensor, which can be regarded as the basic
equation governing integrated photoelasticity. In section~\ref{CharaDirections} we
introduce our new scheme for determining characteristic parameters for a given
optical transfer matrix in terms of an eigenvalue problem. In
section~\ref{PoincareEqTheorem} we explain the relation of the results so obtained
with the Poincar\'e Equivalence Theorem, which was discussed more thoroughly in
Ref.~\cite{HammerJModOpt2004a}; and we define the equivalent optical model, and the
associated characteristic parameters, in terms of the characteristic directions
determined by the method introduced in section~\ref{CharaDirections}. In
section~\ref{Summary} we summarize the paper.

\section{The equations governing integrated photoelasticity}
\label{Equations}

We consider a non-magnetic material which is transparent for optical wavelengths; and
which is optically isotropic ($\epsilon_{ij} = \epsilon \delta_{ij}$) when unloaded
but becomes birefringent when under external load. Plastics and certain resins
exhibit these properties.

As explained in Ref.~\cite{HammerEA-JOSAA-2005a}, the spatial variation of the
optical inhomogeneity in the material---acquired under loading---can be characterised
by the trace $(1/3) \tr \epsilon$ of the dielectric tensor; a heuristic length scale
$l_0$ then gives the distance over which the relative change in $(1/3) \tr \epsilon$
is comparable to one. Furthermore, since the loaded medium is optically anisotropic
at each point, two preferred polarization
directions\cite{BornWolf,Fowles,SommerfeldOptics,Ditchburn,Longhurst} exist for each
direction of propagation. If the medium were homogeneously anisotropic, like a
crystal, these preferred directions would be constant through the whole medium; in
photoelasticity, however, the anisotropy in $\epsilon_{ij}$ will vary at each point
in the material, so that another heuristic scale $l_p$, denoting the distance along
which the relative change of a given preferred polarization direction is comparable
to one, may be defined. If these two scales are substantially larger than the
wavelength of the light passing through the material, $l_0, l_p \gg \lambda$, the
propagation of light through the specimen can be described in the {\it
geometrical-optical approximation}~\cite{BornWolf,SommerfeldOptics}, where the
complex electric field of the light beam may be written in the form
\begin{equation}
\label{geomopt1}
 \wh{\v{E}}(\v{x},t) = \v{E}(\v{x})\, e^{i\phi(\v{x}) - i\omega t}
 \quad.
\end{equation}
Here, the {\it eikonal} $\phi(\v{x})$ describes a locally-plane wave, with local wave
vector $\na \phi(\v{x})$, and local amplitude $\v{E}(\v{x})$. By assumption, both of
these quantities vary weakly on the length scale of a wavelength $\lambda$. It is
then convenient to introduce a dimensionless scale~\cite{FKN},
\begin{equation}
\label{scale3}
 \alpha = \max \left\{ \frac{\lambda}{l_0},\, \frac{\lambda}{l_p}
 \right\} \quad,
\end{equation}
in terms of which the geometrical-optical limit is simply characterised by $\alpha \ll
1$.

Furthermore, an explicit measure of anisotropy may be given by the relative deviation
of $\epsilon$ from its isotropic value~\cite{HammerEA-JOSAA-2005a},
\begin{subequations}
\label{scale5}
\begin{align}
 A_{ij} & = \frac{ \epsilon_{ij} - \delta_{ij} \epsilon}{\epsilon}
 \quad, \label{scale5a} \\
 \beta & \sim \max \left\| A_{ij} \right\| \quad, \label{scale5b}
\end{align}
\end{subequations}
where the global maximum $\beta$ characterises the magnitude of anisotropy in the
dielectric tensor. In principle, this inhomogeneous anisotropy would lead to a
continuous splitting of light rays in the medium~\cite{FKN}, due to the fact that two
distinct phase velocities, and two distinct ray velocities, for any given propagation
direction exist at each point~\cite{BornWolf,SommerfeldOptics,Longhurst}. However,
experimentally, no ray splitting is seen in typical photoelastic measurements;
rather, light rays propagate, for all practical purposes, along straight lines
through the object, and the only effect of the optical tensors on the light beam is
to rotate the preferred polarization directions. This observational fact indicates
that, in the context of (industrially relevant) photoelasticity, the anisotropy of
$\epsilon_{ij}$ is so weak as to permit a replacement of the two actual rays by one
single, ``effective'' ray, which is obtained from the isotropic part $(1/3) \tr
\epsilon$ of the dielectric tensor alone. The mathematical condition for this to be
true has been given in Ref.~\cite{FKN},
\begin{equation}
\label{elec5}
 \frac{\beta}{\alpha } \lesssim 1 \quad.
\end{equation}
In this case, the local wave vector $\na \phi$ in eq.~(\ref{geomopt1}) may be taken
as globally constant, $\na \phi = \v{k}$, so that~(\ref{geomopt1}) describes a
strictly plane wave
\begin{equation}
\label{geomopt2}
 \wh{\v{E}}(\v{x},t) = \v{E}(\v{x})\, e^{i \v{k} \cdot \v{x} - i\omega
 t} \quad.
\end{equation}
Only the weakly varying amplitude $\v{E}(\v{x})$ then encodes the structural
inhomogeneities in the material.

In the limit of the geometrical-optical approximation $\alpha \ll 1$, and under the
condition of negligible ray splitting~(\ref{elec5}), the condition of {\it weak
anisotropy} $\beta \ll 1$ is automatically satisfied. In this case the electric field
may be taken as transverse to the propagation direction $\bm{\kappa} = \v{k} / k$ of
the light beam~\cite{HammerEA-JOSAA-2005a}. Upon inserting the trial
solution~(\ref{geomopt2}) into the wave equation
\begin{equation}
\label{wave1}
 \Delta \v{E} - \mu_0 \underline{\epsilon} \ddot{\v{E}} = 0 \quad,
\end{equation}
and retaining only first-order spatial derivatives---corresponding to the
geometrical-optical limit---we arrive at an equation of the form
\begin{equation}
\label{poltransfer1}
 \bm{\kappa} \times \Big( \bm{\kappa} \times \v{E} \Big) - \frac{i}{k}
 \Big\{ \na \times \big( \bm{\kappa} \times \v{E} \big) + \bm{\kappa}
 \times \big( \na \times \v{E} \big) \Big\} + \mu_0 u^2
 \underline{\epsilon} \v{E} = 0 \quad.
\end{equation}
Here, $u = \omega/k$ is the phase velocity in the unloaded material, and
$\underline{\epsilon}$ is the dielectric tensor. The longitudinal component of
eq.~(\ref{poltransfer1}), obtained by projection onto the unit vector $\bm{\kappa}$,
can be neglected in the geometrical-optical limit. In a coordinate system for which
$\bm{\kappa}$ is along the $z$ axis, eq.~(\ref{poltransfer1}) becomes
\begin{equation}
\label{elec16}
\frac{d}{dz} \left[ \begin{mat}{1} & E_1 \\ & E_2 \end{mat} \right] =
i \frac{\pi}{\lambda} \left[ \begin{mat}{2} & A_{11} && A_{12} \\ &
A_{21} && A_{22} \end{mat} \right] \left[ \begin{mat}{1} & E_1 \\ &
E_2 \end{mat} \right] \quad,
\end{equation}
with $A_{ij}$ as defined in eq.~(\ref{scale5a}). The solution of~(\ref{elec16}) can
be expressed in terms of a transfer matrix $U$ such that
\begin{equation}
\label{elec17}
 \left[ \begin{mat}{1} & E_1(z) \\ & E_2(z) \end{mat} \right] =
 U(z,z_0) \left[ \begin{mat}{1} & E_1(z_0) \\ & E_2(z_0) \end{mat}
 \right] \quad,
\end{equation}
where $U$ satisfies
\begin{equation}
\label{elec18}
\begin{aligned}
 \frac{d}{dz} U(z,z_0) & = i \frac{\pi}{\lambda}\, A_{\bot}(z)\,
 U(z,z_0) \quad, \\[10pt]
 U(z_0,z_0) & = \Eins_2 = \left( \begin{mat}{2} & 1 && 0 \\ & 0 && 1
 \end{mat} \right)  \quad.
\end{aligned}
\end{equation}
The system~(\ref{elec18}) is equivalent to an integral equation
\begin{equation}
\label{forward}
 U(z,z_0) = \Eins_2 + i \frac{\pi}{\lambda} \int\limits_{z_0}^{z}
 dz_1\; A_{\bot}(z_1)\, U(z_1, z_0) \quad,
\end{equation}
where $A_{\bot}$ denotes the matrix of transverse components of $A_{ij}$ as they
appear in eq.~(\ref{elec16}), and $\lambda$ is the wavelength in the unloaded
material. Eq.~(\ref{forward}) is the basic equation governing integrated
photoelasticity: it determines the transfer matrices $U$ for a given light ray, as
functions of the anisotropic part $A_{ij}$ of the dielectric tensor in the interior
of the medium.  It can be shown that $U$ must be unitary, preserving the norm of the
complex electric field vector. This is clear on physical grounds, as preservation of
the norm of $\v{E}$ implies preservation of intensity, so unitarity here just
expresses energy conservation of the light beam. This is just to be expected, since
we have assumed a non-absorbing medium.

Eq.~(\ref{forward}) is manifestly non-linear in $U$, since the transfer matrices
appear under the integral on the right-hand side. This can be seen by iteratively
inserting the left-hand side of eq.~(\ref{forward}) into the right-hand side, leading
to the Born-Neumann series
\begin{equation}
\label{elec20}
 U(z,z_0) = \Eins_2 + \left( i \frac{\pi}{\lambda} \right)
 \int\limits_{z_0}^z dz_1\; A_{\bot}(z_1) + \left( i
 \frac{\pi}{\lambda} \right)^2 \int\limits_{z_0}^z dz_1\;
 A_{\bot}(z_1) \int\limits_{z_0}^{z_1} dz_2\; A_{\bot}(z_2) + \cdots
\end{equation}
This nonlinearity provides one of the major mathematical challenges in the theory of
integrated photoelasticity. As a consequence, the inverse problem of determining
$A_{ij}$ in terms of a collection of transfer matrices is highly non-trivial.

\section{Characteristic directions of linear optical elements}
\label{CharaDirections}

In the previous section we have seen that, in the theory of integrated
photoelasticity, a transfer matrix $U$ can be assigned to each light ray which passes
a photoelastic material through a given point, and along a specified direction. This
means that the associated light path may be regarded as a linear lossless optical
element which acts on the given light ray by changing its polarization form. As
explained in the introduction, there are several occasions in photoelastic stress
analysis, hybrid FEM-photoelastic methods, and methods determining the numerical
stability of numerical algorithms aiming at reconstructing the dielectric tensor from
the collection of---tomographically determined---transfer matrices, where the
decomposition of a given transfer matrix $U$ according to the equivalent optical
model is an issue. As will be shown in this section, the major step in this task is
to determine the characteristic directions of the transfer matrix $U$.

As shown in~\cite{HammerJModOpt2004a}, a lossless linear passive optical device
possesses in general two so-called {\it primary characteristic
directions}~\cite{Aben1966a} $w_m = (\cos\gamma_m, \sin\gamma_m), m=1,2$, in the
plane perpendicular to the entry of the optical element, which have the following
significance: if a light beam at the {\it entry} is plane-polarized in one of the
directions $w_m$ (our convention is such as to define the direction of polarization
along the electric displacement field $\mathbf{D}$) it will leave the device again in
a state of plane polarization, with the plane oriented along unit vectors $w'_m =
(\cos \gamma'_m, \sin \gamma'_m), m=1,2$, called the {\it secondary characteristic
directions}. In contrast, for any direction other than $w_{1,2}$ the beam at exit
will in general be elliptically polarized. The two primary as well as the two
secondary characteristic directions are always perpendicular to each other, so that
it suffices to specify the angle $\gamma$ and $\gamma'$ of the first elements $w_1 =
(\cos \gamma, \sin\gamma)$ and $w'_1 = (\cos \gamma', \sin \gamma')$, respectively;
the second element $w_2, w'_2$ is then determined up to a sign. Since the
polarization state at the exit is again linear the optical device, represented by the
unitary matrix $U$, must act on the real polarization vector $w_m$ according to
\begin{equation}
\label{action1}
 U\, w_m = e^{i \Phi_m}\, w'_m \quad,
\end{equation}
where $\Phi_1, \Phi_2$ are the phases picked up by the light beam
entering along $w_1, w_2$, respectively. Our goal is to determine a
consistent choice of primary and secondary characteristic vectors,
together with appropriate values for the phases $\Phi_m$, from a given
transfer matrix $U$.

Since $U$ is unitary, its determinant is a unimodular number
\begin{equation}
\label{action10101}
 \exp(2i\phi) = \det U \quad,
\end{equation}
hence we can factorize $U$ into
\begin{equation}
\label{action101}
 U = \exp(i\phi)\, S \quad,
\end{equation}
where $S$ is now a unimodular unitary matrix, $\det S = 1$. The choice of $S$ is not
unique, since both $S$ and $-S$ satisfy $\det S =1$. The phase $\phi$ can be computed
from~(\ref{action10101}) modulo $\pi$, the ambiguity in sign obviously related to the
double-valuedness $\pm S$ of the $SU(2)$ factor. We therefore need to stipulate an
explicit convention for the two possibilities in the factorization: we choose $\phi$
to be the {\it smallest possible non-negative} solution of~(\ref{action10101}). Then
$S$ is uniquely determined by eq.~(\ref{action101}).

We can now rewrite~(\ref{action1}) as
\begin{equation}
\label{action103}
 S\, w_m = e^{i\Phi'_m }\, w'_m \quad, \quad \Phi'_m = \Phi_m - \phi
 \quad.
\end{equation}
In principle we can determine the angles $\gamma$ and $\gamma'$ of the primary and
secondary directions $w_m$ and $w'_m$ by parametrising the manifold of $SU(2)$
matrices $S$ in a suitable way and then using elementary trigonometric relations to
express these angles in terms of the coordinates on the $SU(2)$ manifold, as was done
in~\cite{HammerJModOpt2004a}. However, a method that does not require an explicit
coordinate chart on the $SU(2)$ manifold will be presented now:

Suppose that eq.~(\ref{action1}) holds. Then~(\ref{action103}) holds as well, and on
taking the complex conjugate of the latter equation we obtain
\begin{equation}
\label{action2}
 S^*\, w_m = e^{-i \Phi'_m}\, w'_m \quad.
\end{equation}
On eliminating $w'_m$ from eqs.~(\ref{action103}) and~(\ref{action2}) we find that
the directions $w_m$ are {\it real} eigenvectors of $S^T S$,
\begin{equation}
\label{action3}
 (S^T S)\, w_m = e^{2i \Phi'}\, w_m \quad,
\end{equation}
where the superscript $T$ denotes a matrix transpose. We therefore need a method to
obtain real eigenvectors from a complex matrix of the form $S^T S$, where $S$ is an
element of $SU(2)$. To this end we show that $S^T S$ commutes with its complex
conjugate $(S^T S)^*$: any $SU(2)$ matrix $S$ can be represented in the well-known
form
\begin{equation}
\label{action4}
 S = \left( \begin{mat}{2} & a && b \\ -&b^* && a^* \end{mat} \right)
 \quad, \quad |a|^2 + |b|^2 = 1 \quad,
\end{equation}
so that a short computation gives
\begin{equation}
\label{action7}
 (S^T S)\, (S^T S)^* = \Big\{ 4\, \Im^2(ab) + \left| a^2 + b^{*2}
 \right|^2 \Big\}\, \Eins_2 \quad.
\end{equation}
Since the right-hand side is real it follows that the left-hand side is equal to its
complex-conjugate; as a consequence, the commutator
\begin{equation}
\label{action8}
 \bigg[ (S^T S), \, (S^T S)^* \bigg] = 0
\end{equation}
must vanish.

This relation is significant, since it can be used, in turn, as a starting point to
determine the characteristic directions in an elegant way: given any linear lossless
device with unitary Jones matrix $U = \exp(i\phi)\, S$, the $SU(2)$ factor $S$ will
satisfy relation~(\ref{action8}). This relation implies that the commutator
\begin{equation}
\label{action9}
\begin{aligned}
 \Big[ \Re S^T S, \, \Im S^T S \Big] & = \left[\, \rule{0pt}{11pt}
 \frac{1}{2} \Big\{ S^T S + (S^T S)^* \Big\}\,, \;\; \frac{1}{2i}
 \Big\{ S^T S - (S^T S)^* \Big\}\, \right] = \\
 & = - \frac{1}{4i} \left[ S^T S, (S^T S)^* \right] = 0
\end{aligned}
\end{equation}
must vanish as well. The real and imaginary parts of $S^T S$ are symmetric, since
$S^T S$ is. As a consequence of the commutativity~(\ref{action9}), both of these
matrices share the same (orthogonal) system of eigenvectors $w_m$ which must be {\it
real} since the real and imaginary parts are so,
\begin{equation}
\label{action10}
 \Re (S^T S)\, w_m = r_m\, w_m \quad, \quad \Im (S^T S)\, w_m = j_m\,
 w_m \quad.
\end{equation}
It follows that $w_m$ are {\it real} eigenvectors of $S^T S$ as well, with
eigenvalues $r_m + i j_m$. On the other hand, since $S^T S$ is unitary, its
eigenvalues must be the unimodular numbers $\exp(2i\Phi'_m)$ appearing in
eq.~(\ref{action3}), so that
\begin{equation}
\label{action11}
 \exp(2i \Phi'_m) = r_m + i j_m \quad.
\end{equation}

The result~(\ref{action10}) shows that the characteristic directions $w_m$ can be
obtained as the {\it real} eigenvectors of the matrices $\Re S^T S$ or $\Im S^T S$
describing the optical element. Its significance lies in the fact that the process of
finding the characteristic directions from a given transfer matrix $U$ becomes
amenable to well-established numerical routines for general eigenvalue problems. This
problem arises e.g. in integrated photoelasticity, or more generally, in any effort
to reconstruct the dielectric tensor inside a transparent but inhomogeneously
anisotropic optical device from tomographic
measurements~\cite{HammerEA-JOSAA-2005a}. -- This outlines the principle of our
method to obtain the characteristic directions of any transfer matrix $U$. To finish
our discussion we now show how to fix the ambiguity in signs of the eigenvectors
appearing in~(\ref{action3}) and~(\ref{action10}), and the associated phase
ambiguity, in a consistent way:

The numerical routine will deliver two eigenvectors $w_m$, but there are four
possible choices
\begin{equation}
\label{ambi1}
 (w_1, w_2) \quad, \quad (w_1, -w_2) \quad, \quad (-w_1, w_2) \quad,
 \quad (-w_1, -w_2) \quad
\end{equation}
for the signs. We thus need to agree on a convention to pick a system
from~(\ref{ambi1}): we first choose from $\pm w_1$ the vector which makes an angle
with the $x$ axis whose modulus does not exceed $\pi/2$, so that the $x$-component of
this vector is always non-negative. Without loss of generality we may assume this to
be true for $+w_1$. Next we choose from $\pm w_2$ that vector which makes the
system~$(w_1, \pm w_2, e_3)$ {\it right-handed}, where it is assumed that the light
beam propagates along $e_3$ towards positive $z$-values. Without loss of generality
we may assume that this is satisfied by $+w_2$. The phases $\Phi'_m$ can now be
obtained from~(\ref{action11}), but obviously only modulo $\pi$,
\begin{equation}
\label{foureigen1}
 \Phi'_1 \quad, \quad \Phi'_1 + \pi \quad, \quad \Phi'_2 \quad, \quad
 \Phi'_2 + \pi \quad.
\end{equation}
Accordingly we can determine the associated secondary characteristic directions
$w'_m$ from~(\ref{action103}), but only up to a sign, due to the phase ambiguity. We
therefore have four possibilities
\begin{equation}
\label{foureigen2}
 (w'_1, w'_2) \quad, \quad (w'_1, -w'_2) \quad, \quad (-w'_1, w'_2)
 \quad, \quad (-w'_1, -w'_2)
\end{equation}
for the secondary system. We now impose two conditions similar to those that made the
choice of $w_m$ unique: firstly, we require that the suitable candidate $\pm w'_1$
for the first element makes an angle with $w_1$ whose modulus is not larger than
$\pi/2$. Assuming that this is the case for $w'_1$, we must have
\begin{equation}
\label{foureigen3}
 w'_1 \cdot w_1 \ge 0 \quad, \quad | \gamma' - \gamma| \le
 \frac{\pi}{2} \quad,
\end{equation}
where $w'_1 = (\cos \gamma', \sin \gamma')$. We then still have the ambiguity of $\pm
w'_2$; our second condition now is to select this sign so that the vector triad
$(w'_1, \pm w'_2, e_3)$ is right-handed; we may assume that $+w'_2$ is the correct
choice.

We have now fixed the ambiguous signs of the characteristic directions; as a
consequence, the phases $\Phi'_m$ are determined by eq.~(\ref{action103}) up to
multiples of $2\pi$. This last indeterminacy is intrinsic and cannot possibly
removed. Thus we stipulate to let the phases $\Phi'_m$ take values in the interval $
- \pi \le \Phi'_m < \pi$, it being understood that the values of $\Phi'_1$ and
$\Phi'_2$ are {\it different}; for, if they were equal, they would have been part of
the phase $\phi$ which was extracted out of $U$ in eq.~(\ref{action10101}). As a
consequence,
\begin{equation}
\label{foureigen5A}
 -2\pi < \Phi'_1 + \Phi'_2 < 2\pi \quad.
\end{equation}

$S$ can now be represented in terms of primary and secondary characteristic
directions, and associated phases, as
\begin{equation}
\label{action1103}
 S = |w'_1\ra\, \exp (i \Phi'_1)\, \la w_1| + |w'_2\ra\, \exp
 (i\Phi'_2)\, \la w_2| \quad,
\end{equation}
where we have denoted (column) vectors as $|w\ra$ and (row) covectors as $\la w|$,
reminiscent to quantum-mechanical conventions.

\section{Relation to the Poincar\'e equivalence theorem}
\label{PoincareEqTheorem}

Finally we show how the representation~(\ref{action1103}) of $S$ in terms of
characteristic directions and phases is related to the decomposition of a lossless
linear optical element according to the Poincar\'e equivalence
theorem~\cite{Poincare1892}: to this end we represent eq.~(\ref{action1103}) in the
basis $e_1, e_2$ of Cartesian coordinate vectors pertaining to the laboratory frame:
using the notation $\la e_i | S| e_j\ra = S_{ij}$ we see that~(\ref{action1103})
takes the form
\begin{equation}
\label{action1105}
 S_{ij} = \sum_{m_1,\, m_2=1}^2 R(-\gamma')_{i m_1}\, J'_{m_1 m_2}\,
 R(\gamma)_{m_2 j} \quad,
\end{equation}
where
\begin{equation}
\label{actrot1}
\begin{aligned}
 R(-\gamma')_{im} & = \la e_i | w'_m \ra \quad, \quad J'_{m_1 m_2} =
 e^{i \Phi'_m} \delta_{m_1 m_2} \quad, \quad R(\gamma)_{mj} = \la
 w_m|e_j\ra \quad, \\
 R(\gamma) & = \left( \begin{array}{@{}*{2}{r@{}c}@{}} & \cos\gamma &
 & \sin\gamma \\ -& \sin\gamma & & \cos\gamma \end{array} \right)
 \quad,
\end{aligned}
\end{equation}
using the notation conventions of~\cite{HammerJModOpt2004a}. The vectors in the pairs
$(w_1, w_2)$ and $(w'_1, w'_2)$ are orthogonal, and have been constructed to make a
right-handed system together with $e_3$. It follows that the matrices $R(\gamma),
R(-\gamma')$ are proper rotation matries having unit determinant, i.e. elements of
$SO(2)$. Then, since $\det S = 1$ it follows that $\det J' = 1$, which implies that
the eigenvalues $\Phi'_m$ must sum up to a multiple of $2\pi$, $\Phi'_1 + \Phi'_2 =
2\pi N$. But, according to~(\ref{foureigen5A}) this restriction can be made stronger,
\begin{equation}
\label{actrot03}
 \Phi'_1 + \Phi'_2 = 0 \quad.
\end{equation}
Finally, on multiplying~(\ref{action1105}) with $\exp(i\phi)$ we find on
using~(\ref{action101}) and~(\ref{actrot03}) that
\begin{subequations}
\label{actrot3}
\begin{align}
 U & = R(-\gamma')\, J(0, \delta)\, R(\gamma) \quad,
 \label{actrot3a} \\
 J(0,\delta) & = \diag\left(\, \exp( -i \delta/2), \, \exp( i
 \delta/2)\, \right) \quad, \quad - \frac{\delta}{2} = \Phi'_1 + \phi
 \quad. \label{actrot3b}
\end{align}
\end{subequations}
We recognize that $J(0,\delta)$ is the Jones matrix of a linear retarder whose fast
axis, for $\delta >0$, coincides with the $x$ axis of the laboratory system, so that
light plane-polarized along $e_1$ ($e_2$) accumulates a relative phase $- \delta/2$
($\delta/2$) on passing through the device, without changing its linear polarization
form, or the orientation of the plane of polarization. On using the fact that the
transfer matrix of a linear retarder with fast axis making a nonvanishing angle
$\gamma$ with the $x$ axis is given by
\begin{equation}
\label{actrot4}
 J(\gamma, \delta) = R(-\gamma) \, J(0,\delta) \, R(\gamma) \quad,
\end{equation}
we can rewrite~(\ref{actrot3a}) in the equivalent forms
\begin{equation}
\label{actrot5}
 U = J(\gamma', \delta)\, R(-\gamma' + \gamma) = R(- \gamma' +
 \gamma)\, J(\gamma, \delta) \quad.
\end{equation}
The decompositions~(\ref{actrot3}),~(\ref{actrot5}) express the fact that any linear
lossless optical device can be replaced by a sequence of one linear retarder and one
or two appropriate rotators, at least as far as its optical properties are
concerned. The fictitious optical device comprised of these retarders and rotators is
called the {\it equivalent optical model}. The three quantities $\gamma, \gamma',
\delta$ are commonly called the {\it characteristic parameters} of the equivalent
optical model~\cite{Aben1966a}, where the angle $-\gamma' + \gamma$ specifying the
equivalent rotator is the same for both forms in eq.~(\ref{actrot5}). In the first
(second) form, $\gamma'$ ($\gamma$) is the angle between the customarily selected
primary characteristic direction---the fast axis of the equivalent retarder---and the
$x$-axis; while $\delta$ is the characteristic phase retardation of the equivalent
retarder in both forms. -- The physical and mathematical content
of~(\ref{actrot3}),~(\ref{actrot5}) is called the {\it Poincar\'e equivalence
theorem}. The decompositions as given above coincide with the forms given
in~\cite{HammerJModOpt2004a}.

\section{Summary} \label{Summary}

We have presented a method to determine the primary and secondary characteristic
directions of a linear lossless optical device from an eigenvalue problem formulated
in terms of the unimodular factor of the transfer matrix of the optical element. This
approach is conceptually more elegant than methods using explicit parametrisations of
the manifold of $SU(2)$ matrices, and is furthermore amenable to pre-implemented
numerical routines, thus making the decomposition of the transfer matrix in terms of
equivalent linear retarders and rotators numerically more convenient. This important
issue arises in the context of stress analysis based on integrated photoelasticity or
hybrid methods combining photoelastic measurements with analytical stress models
and/or numerical Finite-Element (FEM) evaluations of the stress tensor field. In
addition, we have given a visual example of how our results can be used to test the
stability of reconstruction algorithms for the dielectric tensor in the interior of a
photoelastic model on noisy measurement data. A brief derivation of the basic
equations governing integrated photoelasticity has been provided. The relation of our
results to the associated Poincar\'e equivalence theorem has been explained.

\section*{Acknowledgements}
 Hanno Hammer wishes to acknowledge acknowledge support from EPSRC grant
 GR/86300/01. -- Hanno Hammer gratefully acknowledges inspiring discussions with
 W. R. B. Lionheart.


\end{document}